\documentclass[apj,usenatbib,twocolumn]{aastex6}

\usepackage{float}
\usepackage{todonotes}
\usepackage{xspace}
\usepackage{graphicx}
\usepackage{subfigure}
\usepackage{amsmath}

\newcommand{\appropto}{\mathrel{\vcenter{
  \offinterlineskip\halign{\hfil$##$\cr
    \propto\cr\noalign{\kern2pt}\sim\cr\noalign{\kern-2pt}}}}}

\def\arcsec{$^{\prime\prime}$}
\def\lya{Ly$\alpha$}

\def\nm{\hbox{nm}}
\def\tmax{T_{max}}
\def\bw90{BW90}

\begin{document} 

   \title{{\large D}esign {\large f}or {\large t}he {\large F}irst {\large N}arrowband {\large F}ilter {\large f}or {\large t}he {\large D}ark 
{\large E}nergy {\large C}amera: {\large O}ptimizing {\large t}he {\large LAGER} {\large S}urvey {\large f}or $z\sim$ 7 {\large G}alaxies}
   
      \author{
         Zhen-Ya\ Zheng \altaffilmark{1},
	James\ E. Rhoads \altaffilmark{2,6}, 
	Junxian\ Wang \altaffilmark{3}, 
	Sangeeta\ Malhotra \altaffilmark{2,6}, 
	Alistair\ Walker \altaffilmark{4}, 
	Thomas\ Mooney \altaffilmark{5}, 
	Chunyan\ Jiang \altaffilmark{1},
	Weida\ Hu \altaffilmark{3}, 
	Pascale\ Hibon\altaffilmark{7}, 					
	Linhua\ Jiang\altaffilmark{8},				
	Leopoldo\ Infante\altaffilmark{9},			
	L. Felipe\ Barrientos\altaffilmark{10},			
	Gaspar\ Galaz\altaffilmark{10},				
	Francisco\ Valdes \altaffilmark{11},			
	William\ Wester \altaffilmark{12}, 			
	Huan\ Yang\altaffilmark{9},				
	Alicia\ Coughlin\altaffilmark{6}, 	 			
	Santosh\ Harish\altaffilmark{6} 	 			
	Wenyong\ Kang\altaffilmark{3},				
	Ali Ahmad\ Khostovan\altaffilmark{6},		
	Xu\ Kong\altaffilmark{3},	 				
	Lucia A.\ Perez\altaffilmark{6},				
	John\ Pharo\altaffilmark{6},				
	Isak\ Wold\altaffilmark{2},	 				
	XianZhong\ Zheng\altaffilmark{13}			
  }
      \altaffiltext{1}{CAS Key Laboratory for Research in Galaxies and Cosmology, Shanghai Astronomical Observatory, Shanghai 200030, China; zhengzy@shao.ac.cn}
      \altaffiltext{2}{ Astrophysics Science Division, Goddard Space Flight Center, 8800 Greenbelt Road, Greenbelt, MD 20771, USA; james.e.rhoads@nasa.gov}
      \altaffiltext{3}{ CAS Key Laboratory for Research in Galaxies and Cosmology, Department of Astronomy, University of Science and Technology of China, Hefei, Anhui 230026, China; jxw@ustc.edu.cn}
      \altaffiltext{4}{ Cerro Tololo Inter-American Observatory, National Optical Astronomy Observatory, Casilla 603, La Serena, Chile}
      \altaffiltext{5}{ Materion Precision Optics, 2 Lyberty Way, Westford, MA 01886, USA}
      \altaffiltext{6}{ School of Earth and Space Exploration, Arizona State University, Tempe, AZ 85287, USA}
      \altaffiltext{7}{ European Southern Observatory, Alonso de Cordova 3107, Casilla 19001, Santiago, Chile}
      \altaffiltext{8}{ The Kavli Institute for Astronomy and Astrophysics, Peking University, Beijing, 100871, China}
      \altaffiltext{9}{ Las Campanas Observatory, Carnegie Observatories}
      \altaffiltext{10}{ Instituto de Astrof\'isica,  Facultad de F\'isica, Pontificia Universidad Cat\'olica de Chile, Santiago, Chile}       
      \altaffiltext{ 11}{ National Optical Astronomy Observatory, 950 N. Cherry Ave., Tucson, AZ 85719, USA}
      \altaffiltext{12 }{ Particle Physics Division, Fermi National Accelerator Laboratory, Batavia, IL 60510, USA}
      \altaffiltext{13 }{ Purple Mountain Observatory, Chinese Academy of Sciences, Nanjing 210008, China}

\begin{abstract}
  We present the design for the first narrowband filter NB964 for the Dark Energy Camera (DECam), 
  which is operated on the 4m Blanco Telescope at the Cerro Tololo Inter-American Observatory.
  The NB964 filter profile is essentially defined by maximizing the power of searching for Lyman 
  alpha emitting galaxies (LAEs) in the epoch of reionization, with the consideration of the night 
  sky background in the near-infrared and the DECam quantum efficiency. The NB964 filter was 
  manufactured by Materion in 2015. It has a central wavelength of 964.2 nm and a full width at 
  half maximum (FWHM) of 9.2 nm. An NB964 survey named LAGER (Lyman Alpha Galaxies in 
  the Epoch of Reionization) has been ongoing since December 2015. Here we report 
  results of lab tests, on-site tests and observations with the NB964 filter. The excellent performances 
  of this filter ensure that the LAGER project is able to detect LAEs at $z \sim$ 7 with a high efficiency.  
\end{abstract}


\section{Introduction}
\label{sec:intro}

The narrowband imaging technique has been widely used to search for high-redshift star-forming galaxies since about two decades ago
 \citep{CowieHu1998,Hu+1998,Rhoads+2000}.
 Targeting the Lyman alpha (\lya) emission produced in high-redshift star formation, the narrow-band imaging can select high-redshift galaxies by the
 excess flux in the narrowband compared to the broadband. 
Thousands of high-redshift Lyman alpha emitting galaxies (LAEs) have been selected this way with ground-based 4-10m telescopes from 
redshift 2 to redshift 6.6 \citep[e.g.,][]{Guaita+2010, Gawiser+2007, Nilsson+2009, Ouchi+2008, Finkelstein+2008, Ouchi+2010, Hu+2010}, 
with a very large fraction being spectroscopically confirmed \citep[e.g.,][]{Kashikawa+2011, Jiang+2017}. 
The large number of LAEs have ensured intense studies of the galaxy formation and evolution \citep[e.g.,][]{Gronwall+2007, Finkelstein+2009, 
Ouchi+2010}, and the large scale structure of the universe  over the cosmic time \citep[e.g.,][]{Steidel+2000, Zheng+2016, Cai+2017, Jiang+2018}.

At $z\gtrsim$ 7, the epoch of reionization (EoR), the resonant scattering of Lyman alpha photons in the neutral hydrogen intergalactic medium (IGM) 
makes it very important to study LAEs in the EoR, which will enable us to put constraints on the fraction of the neutral hydrogen in the EoR and its 
evolution with time\citep[][]{MalhotraRhoads2004}. 
 However, previous searches at $z\gtrsim$ 7 only yielded about two dozen LAEs
 \citep{WillisCourbin2005, Iye+2006, Cuby+2007, Ota+2008, Ota+2010, Hibon+2011, Hibon+2012, Shibuya+2012, Konno+2014}, 
 with several being spectroscopically confirmed. 
The small number of $z\gtrsim$ 7 LAEs were mainly limited by the moderate field-of-views (FOVs, $\sim$ 0.2 deg$^2$) of the telescopes/instruments 
with which the surveys were carried out.

Recently, thanks to the emergence of the new-generation survey cameras with very large FOVs, e.g., Dark Energy Camera \citep[DECam, FOV $\sim$ 3 deg$^2$,][]{Flaugher+2015} and 
Hyper-Suprime Camera \citep[HSC, FOV $\sim$ 1.8 deg$^2$,][]{Miyazaki+2012}, which take advantage of the full prime focal plane of the telescopes,
we can carry out narrowband imaging surveys for LAEs in the EoR with $\gtrsim 10$ times higher efficiency than previous ones. 
Two such narrowband surveys with DECam \citep{Zheng+2017} and HSC \citep{Ouchi+2018, Itoh+2018} have been underway. 
They use narrowband filters designed to avoid strong sky emission and absorption in the bandpass. 
However, since such narrowband filters are quite large (e.g., filter diameter $\geq$ 600mm) and operated with a much smaller focal ratio 
(about f/3 to f/2 for the prime focal plane), the design and manufacture of these narrowband filters are quite challenging. 
The narrowband filter encounters wavelength shifts in the case of small focal ratios, and the wavelength shifts may vary at different locations 
of the filter. This could bring unwanted strong OH emission lines into the filter, and decrease the survey efficiency for $z\geq$ 7 galaxies.

In this paper, we discuss the design for the first narrowband filter for DECam, the NB964 filter. 
With this filter, a survey named LAGER (Lyman Alpha Galaxies in the Epoch of Reionization)
is ongoing since December 2015 \citep[see ][ for the first results of LAGER]{Zheng+2017,Hu+2017}. 
The parameters of the narrowband filter NB964 are optimized to maximize the survey efficiency for $z\sim$ 7 galaxies.

This paper is organized as follows. In Section \ref{sec:method}, we optimize the design for the narrowband filter NB964 for DECam.
In Section \ref{sec:made}, we introduce the production of the narrowband filter NB964. In Section \ref{sec:test}, we report  
lab tests, on-site tests and observations with the narrowband filter NB964, and discuss the impact of the filter profile. 
We finally summarize the results in Section \ref{sec:conclusion}.

\begin{figure}[t]
\centering
\includegraphics[width=0.45\textwidth]{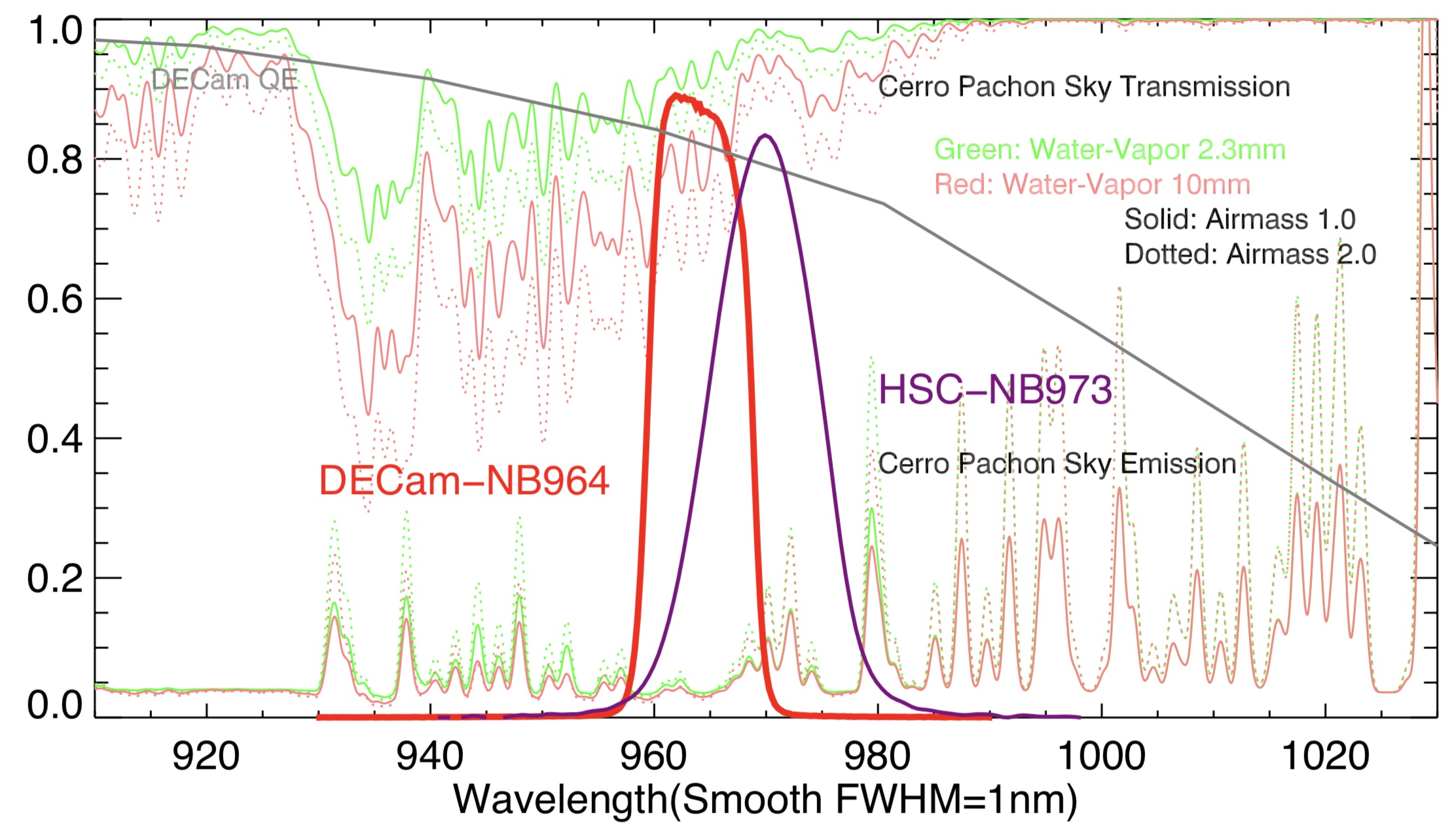}
\caption{Sky transmission and emission in the wavelength range of 910-1030 nm, overlaid with
the DECam NB964 transmission profile (in f/2.9, thick red curve), the HSC NB973 profile (in f/2.0, 
purple curve) and the DECam quantum efficiency (QE, grey curve).
The sky transmission and emission (smoothed with 1 nm resolution)  are plotted with 
two airmasses (1.0 and 2.0) and two water-vapor columns (2.3 mm and 10 mm), respectively.
The unit of the sky emission is arbitrary. 
}
\label{fig:fig1}
\end{figure}

\section{Optimizing the narrowband filter NB964 for DECam} \label{sec:method}

We design the narrowband filter with the purpose of optimizing the narrowband survey for $z\sim$ 7 \lya\ galaxies.
The filter bandpass, defined by the central wavelength $W_c$ and its full width at half maximum FWHM, is specified by considering
the night sky OH emission spectrum, the H$_2$O absorption spectrum, 
and the DECam quantum efficiency (QE) curve close to 1$\mu m$. We also take into account the effects of different angles of 
incidences (AOIs) and focal ratios (F-ratios) on the filter design. Finally, the filter parameters are well tuned 
based on the \lya\ luminosity function model, so that the survey efficiency for LAEs at $z\sim$ 7 is maximized.

\subsection{CCD QE and Sky-background close to 1$\mu m$}
\label{sec:method:sky}

As a ground-based optical survey with CCDs, the CCD QE curve and the sky-background spectra (Fig. \ref{fig:fig1}) can roughly determine the 
wavelength range of the narrowband filter, and hence set the redshift range of the corresponding high-$z$ LAE survey. 
The CCD QE drops sharply towards 1$\mu m$, so that the central wavelength 
of the filter should be set shorter than 1 $\mu m$, which gives an upper boundary of the LAE surveys' redshift of
$z\lesssim$7.3. 

We employ the near-Infrared sky background data from Gemini South\footnote{Gemini South is located at Cerro Pachon, Chile. 
The observational conditions of the site are listed at \url{https://www.gemini.edu/sciops/telescopes-and-sites/observing-condition-constraints}}
 to investigate the sky background at wavelength around 970nm, which is the wavelength of the $z\sim$ 7 \lya\ line in the observer-frame. 
 The Gemini South Observatory is located on a mountain adjacent to Cerro Tololo where DECam resides, only about 10 km away but at 500m greater altitude.
 Therefore, it would be a good approximation to use the sky background data at the site of Gemini South to design our narrowband filter. 
 
 The Gemini website lists the near-Infrared sky transmission and emission data under airmass values of 1.0, 1.5, 2.0, and water-vapor 
 column values of 2.3 mm, 4.3 mm, 7.6 mm, and 10 mm, respectively. 
 Both airmass and water-vapor column would affect sky transmission 
 and emission. A larger column water-vapor means larger absorption on the sky transmission, and a higher
 airmass leads to stronger sky emission. In the following, we choose a water-vapor column of 10 mm which is the worst case, 
 and the airmass of 1.0 and 2.0 to examine the filter bandpass.

\begin{figure}[t]
\centering
\includegraphics[width=0.45\textwidth]{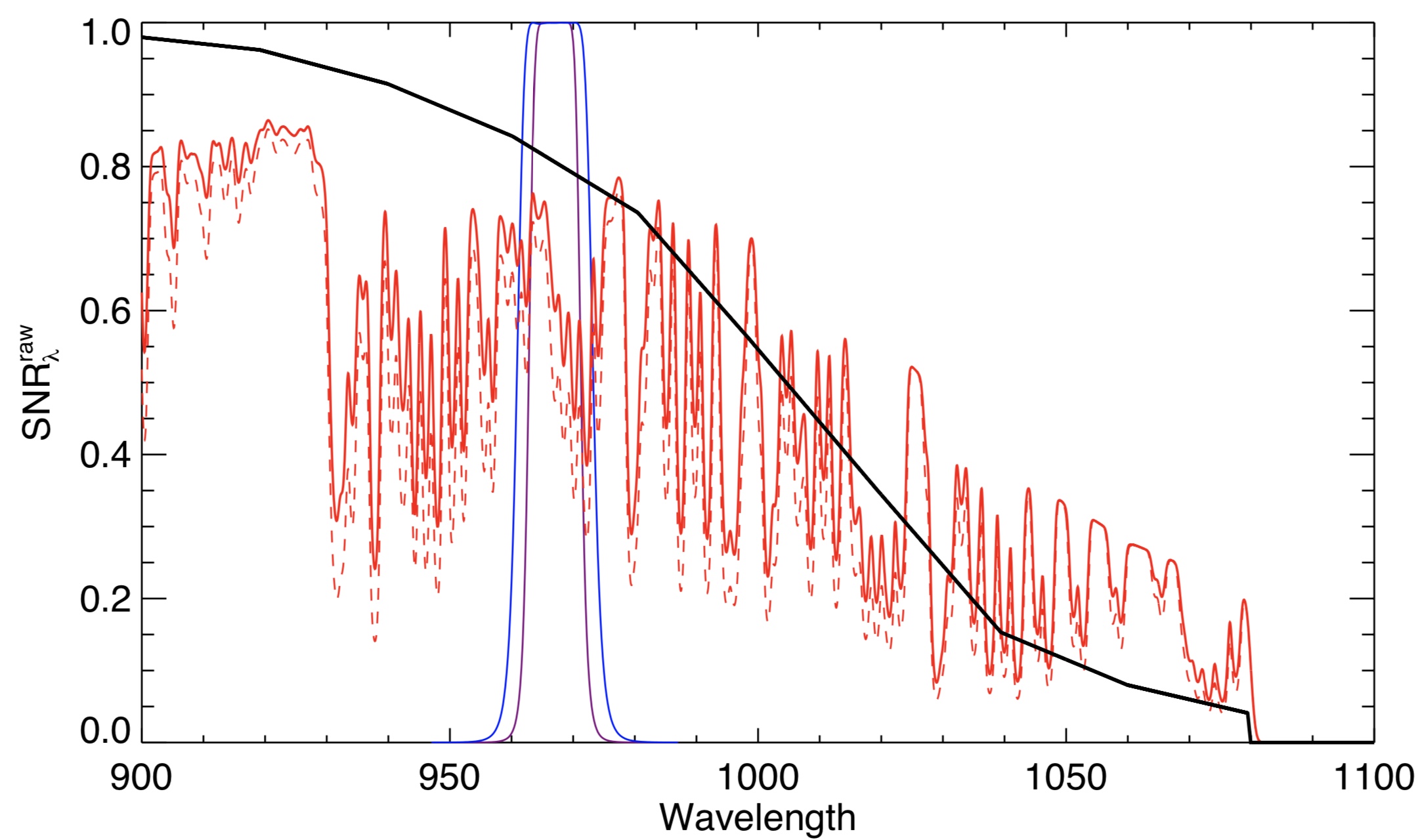}
\caption{SNR$^{\textbf{raw}}_{\lambda}$ spectra calculated from Eq. \ref{eq1} 
with an airmass of 1.0 (solid red) and 2.0 (dashed red), respectively, both with water-vapor column of 10mm. 
Both spectra are smoothed with 1nm resolution for display. The DECam QE (black curve) is overplotted. For comparison, we also plot  
two models of the filter profile generated by Materion, with FWHM = 8.74 nm (purple) and FWHM = 12.76 nm (blue), respectively. }
\label{fig:fig2}
\end{figure}

\subsection{Optimizing the Narrowband Filter Profile } \label{sec:method:optimizing}
In this section, we discuss how to design the NB964 profile for the purpose of maximizing the LAE survey capacity
with the sky background and CCD QE close to 1$\mu m$ shown in Sec. \ref{sec:method:sky}. We also consider the changes of the filter profile modified 
by different AOIs and F-ratios. 

\subsubsection{SNR Spectrum from Sky-background and CCD QE}\label{sec:method:optimizing:SNR}
In the case of sky dominated background, the signals and the noises of a narrowband imaging survey are affected
 by the sky transmission and the sky emission, respectively. Therefore, 
we can define a raw signal-to-noise ratio (SNR) spectrum, which is a ratio of 
source signal to the sky background fluctuation (Sky\_BKG\_RMS) as a function of wavelength $\lambda$,  to estimate the approximate filter bandpass. 
 The raw SNR spectrum is defined as
\begin{equation}\label{eq1}
\textrm{SNR}^{\textbf{raw}}_{\lambda} =  \frac{\textrm{Source\_Signal}}{\textrm{Sky\_BKG\_RMS}} = \frac{T_\textrm{Exp} O_\lambda T_\lambda \textrm{QE}_\lambda}{\sqrt{T_\textrm{Exp} \textrm{Sky}_\lambda\textrm{QE}_\lambda}},
\end{equation}
where T$_\textrm{Exp}$ is the survey exposure time with the narrowband filter, $O_\lambda$ is the input source spectrum, $T_\lambda$ is the sky transmission, 
Sky$_\lambda$ is the sky emission spectrum, and QE$_\lambda$ is the DECam quantum efficiency. 
Assuming a flat input spectrum ($\propto \lambda^{-2}$), we obtain a SNR$^{\textbf{raw}}_{\lambda}$ spectrum as shown in Fig. \ref{fig:fig2}. 
The SNR$^{\textbf{raw}}_{\lambda}$ spectrum follows the CCD QE curve on the whole, while in detail the 
SNR$^{\textbf{raw}}_{\lambda}$ spectrum is modified by the sky transmission and emission. 
In the wavelength range of 930 nm $< \lambda < $ 1$\mu m$, there are two 
SNR$^{\textbf{raw}}_{\lambda}$ peaks at $\sim$962nm and $\sim$975nm, separated by a big gap in between. 
Therefore, an efficient narrowband filter for LAEs at $z\gtrsim$ 7 should have the central wavelength set at around one 
of the two peaks. To display the possible location of the filter in the raw SNR spectrum, in Fig. \ref{fig:fig2} we overplot two models 
of the filter profile generated by Materion around the first peak, with FWHM = 8.74 nm and 12.76 nm, respectively.

    \begin{figure}[t]
   \begin{center}
   \includegraphics[width=0.45\textwidth]{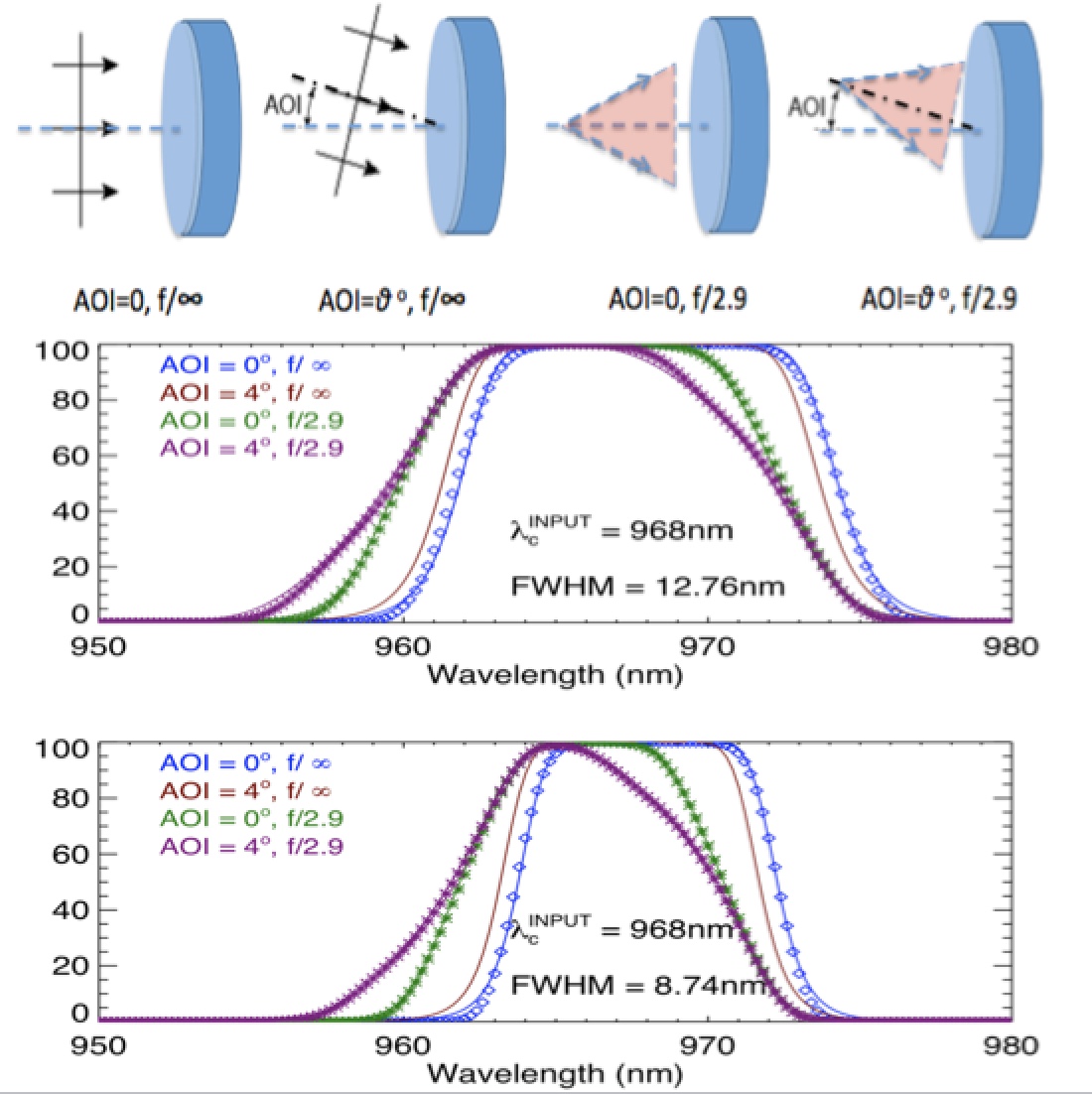}
   \end{center}
   \caption[example] 
{ \label{fig:fig3} 
Effects of AOI and F-ratio on the filter profile. {\it Top:} four different conditions of AOI and F-ratio (AOI = [0$^\circ$, 4$^\circ$] and F-ratio = [$f/\infty$, $f/2.9$]). 
{\it Middle:} model profiles of the narrowband filter under these conditions of AOI and F-ratio with a filter width of FWHM = 8.74 nm. {\it Bottom:} 
same as above, but with a filter width of FWHM = 12.76 nm. The detailed calculations of model profiles are explained in \S \ref{sec:method:optimizing:Fratio}.
}
\end{figure}

 \subsubsection{Effects of AOI and F-ratio on the Filter Profile}\label{sec:method:optimizing:Fratio}
 In this section, we create the model filters to see the effects of AOI and F-ratio on the filter bandpass. 
 AOI is the light beam's incidence angle, and F-ratio is the ratio of an optical system's focal length to 
 the diameter of the entrance pupil. {There are several F-ratios for DECam and the Blanco 4-m telescope. 
 For the prime focus of the Blanco 4-m telescope, the focal ratio is f/2.7 \citep[][]{Flaugher+2015}. With the optical 
 design, the focal ratio for the DECam focal plane is f/3.1\footnote{See DECam Data Handbook at \url{https://www.noao.edu/meetings/decam/media/DECam_Data_Handbook.pdf}}.
 While at the position of the filter slot, the focal ratio is f/2.9 for NB964's operational environment. }
Ideally the filter profile is assumed in a square "boxcar" shape, which is adopted in the previous section. 
In reality, the manufactured filter usually has a shape $\Psi_\lambda(W_c,\textrm{FWHM})$ between a square "boxcar" and a gaussian, 
and the filter profile changes at different AOI and F-ratio. 

We discuss the changes of the filter profile in 4 cases of different AOI and F-ratio, which is presented in Fig. \ref{fig:fig3}.
In the case of a collimated light beam falling at normal incidence, which means AOI $\theta$ = 0$^\circ$ and F-ratio f/$\infty$, 
the model profile of a narrowband filter, $\Psi^0_\lambda$, can be fitted with a convolution of a 
square "boxcar" shape (width = FWHM) and a gaussian function ($\sigma$ = FWHM/2.35).
If a collimated light beam falls at a non-zero incidence angle to the filter, i.e., AOI $\theta\neq$ 0$^\circ$ and F-ratio f/$\infty$, 
the bandpass of the interference filter shifts towards bluer wavelengths, according to 
\begin{eqnarray}\label{eq2}
\Delta\lambda & = & \lambda - \lambda_0 \\ \nonumber
 & = & \lambda_0 \times (\sqrt{1-(\frac{\textrm{sin}\theta}{n})^2} -1 )  \\ \nonumber
 & \simeq & -\lambda_0 \times \frac{(\textrm{sin}\theta)^2}{2n^2} \simeq - \frac{\lambda_0\theta^2}{2n^2}, \textrm{  if }\theta \ll \textrm{1,}
 \end{eqnarray}
where $\lambda_0$ is the wavelength of the input light and $n\sim$ 2.0 is the mean refractive index of the material of multilayers. 
For our narrowband filter, $\lambda_0$ approximately equals to 970 nm.
Therefore, when the incidence angle $\theta$ of the input light beam varies from 0$^\circ$ to 4$^\circ$, the modeled filter profile would shift
from 0 nm to -0.59 nm in the wavelength direction entirely. 

If the input light beam is not collimated, e.g., in the case of $f/2.9$ and zero incidence angle, 
 the input angle $\theta$ would vary from 0$^\circ$ at the filter center to 9.88$^\circ$ at the filter edge. 
 At the filter's annulus $r$ to $r+dr$, which corresponds to the input angle $\theta$ to $\theta + d\theta$, 
 the wavelength shift of the transmission curve is a value between $\Delta\lambda$ and $\Delta\lambda$ + $d\Delta\lambda$.
  Based on Eq. \ref{eq2}, the wavelength shifts $\Delta\lambda$ vary from 0 nm in the filter center to -3.6 nm on the edge of the filter.
  The corresponding normalization of the transmission curve in that annulus is scaled by the ratio of the solid angle of the annulus to that of the whole aperture.
  Therefore, the total output of the filter profile $\Psi_\lambda(W_c,\textrm{FWHM})$ is an integration over the aperture of the filter radially, 
 which is
 \begin{eqnarray}\label{eq2a}
\Psi_\lambda(W_c&,&\textrm{FWHM})  =   \int_0^{\frac{9.88^\circ}{180^\circ}\pi}\frac{\Delta S_\theta}{S} \Psi^0_{\lambda+d(\Delta\lambda_\theta)}d\theta, \\ \nonumber
 & = &  \int_0^{\frac{9.88^\circ}{180^\circ}\pi}\frac{((\theta+d\theta)^2 - \theta^2)}{(\pi\times9.88^\circ/180^\circ)^2} \Psi^0_{\lambda+d(\Delta\lambda_\theta)}, \\ \nonumber
 & = &  \int_0^{\frac{9.88^\circ}{180^\circ}\pi}\frac{2\theta d\theta}{(\pi\times9.88^\circ/180^\circ)^2} \Psi^0_{\lambda+d(\Delta\lambda_\theta)}, \\ \nonumber
 & \approx & \int_0^{-3.6 nm}\frac{2n^2}{(\pi\times9.88/180)^2\lambda_0} \Psi^0_{\lambda+d(\Delta\lambda)} d(\Delta\lambda).
 \end{eqnarray}
 This equation describes a convolution of the normal-incidence filter profile with a boxcar function that extends from the maximum wavelength shift to zero wavelength shift.
 The output profile is shown as the green lines in Fig. \ref{fig:fig3}.
 If the input f/2.9 light beam falls at a non-zero incidence angle, the filter profile will be distorted in a way that  
 a part of the profile at the red shoulder will be cut and added back to the profile at the blue wing, which is shown as the purple lines in Fig. \ref{fig:fig3}.
 
 In the following sections, we define the filter bandpass in the cases of filter profiles with F-ratio of f/2.9 and two AOI values of 0 deg and 4 deg, respectively.

 \subsubsection{Filter Design for the Maximization of the LAE Survey}
 \label{sec:method:optimizing:survey}

 We define the real filter bandpass with which the LAE survey efficiency is maximized. 
Including the real filter shape $\Psi_\lambda(W_c,\textrm{FWHM})$, the raw SNR spectrum in Eq. \ref{eq1} is updated to be 
\begin{equation}\label{Eq3}
\textrm{SNR}_{\lambda}(W_c, \textrm{FWHM}) = \frac{T_\textrm{Exp} \Psi_\lambda(W_c,\textrm{FWHM})O_\lambda T_\lambda \textrm{QE}_\lambda}{\sqrt{T_\textrm{Exp} \Psi_\lambda(W_c,\textrm{FWHM})\textrm{Sky}_\lambda \textrm{QE}_\lambda}},
\end{equation}
where the definitions are the same in Eq. \ref{eq1} except that the filter profile $\Psi_\lambda(W_c,\textrm{FWHM})$ is added. 
The central wavelength $W_c$ and $\textrm{FWHM}$ should be chosen to avoid the strong sky emission lines inside, so as to 
maximize the detecting efficiency for LAEs.

\begin{figure}[t]
   \begin{center}
      \includegraphics[width=0.5\textwidth,angle=0]{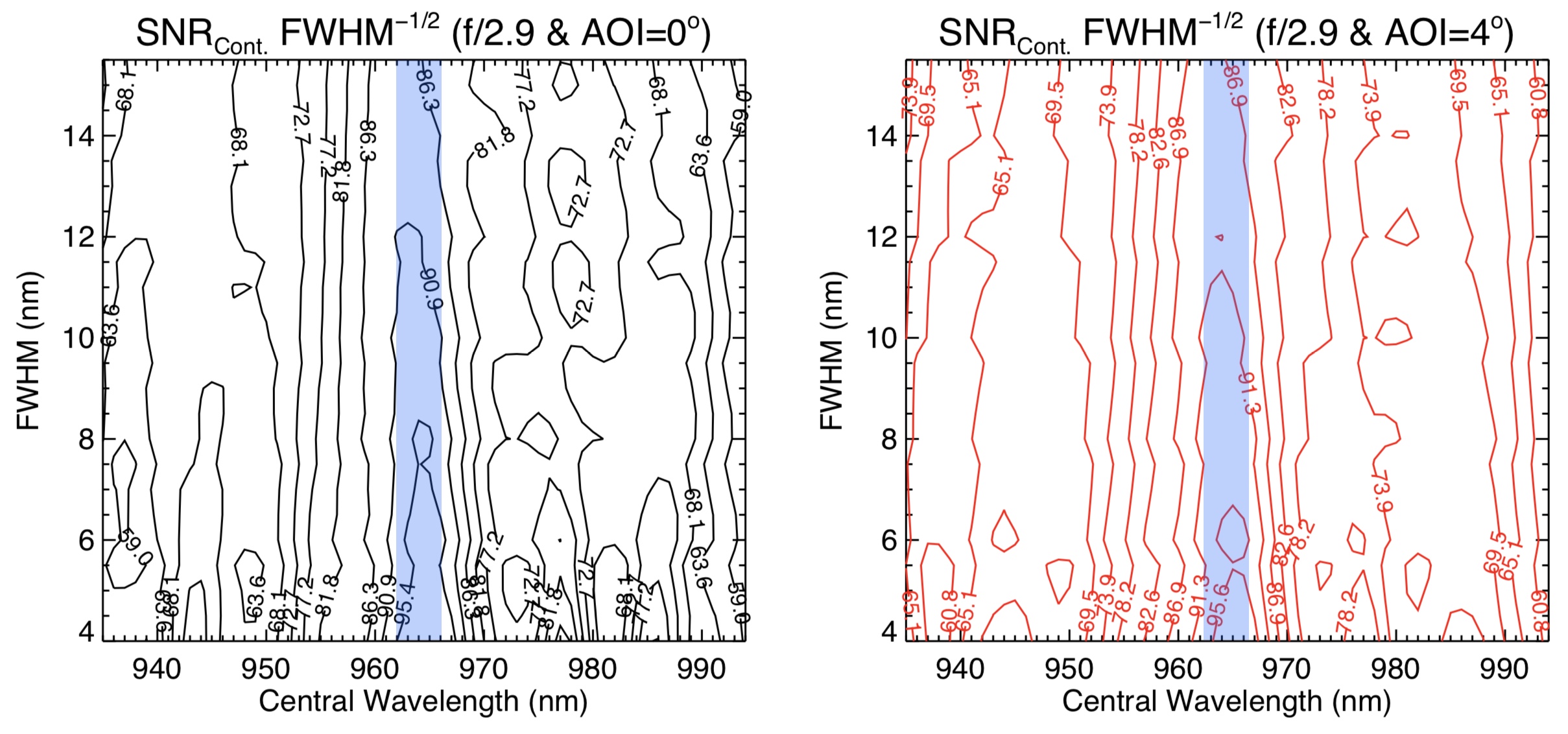} 
   \end{center}
   \caption[example] 
{ \label{fig:fig4} Contours of SNR$_{\textrm{Cont.}} \textrm{FWHM}^{-1/2}$ as a function of the central wavelength $W_c$ and FWHM in the conditions of f/2.9 \& AOI=0$^o$ (left panel, black contours) and f/2.9 \& AOI=4$^o$ (right panel, red contours). The shaded blue regions show that narrowband profiles with a central wavelength of $\sim$964$\pm$2 nm would have best SNR values. 
Contours of SNR$_{\textrm{Line}} \textrm{FWHM}^{1/2}$ show a consistent results. 
}
\end{figure}

Considering the two cases of the input spectrum $O(\lambda)$, a line and a continuum,
the total SNR (integrated over the filter profile) would be SNR$_\textrm{Line}$ $\appropto$ $\textrm{FWHM}^{-1/2}$ 
 and SNR$_\textrm{Cont.}$ $\appropto$ $\textrm{FWHM}^{1/2}$ for the line and the continuum, respectively.
Therefore, the distributions of SNR$_{\textrm{Line}} \times \textrm{FWHM}^{1/2}$ and SNR$_{\textrm{Cont.}} \times \textrm{FWHM}^{-1/2}$
are a function of the central wavelength $W_c$ and roughly independent of FWHM. We display the 
distributions of SNR$_{\textrm{Line}} \times \textrm{FWHM}^{1/2}$ and SNR$_{\textrm{Cont.}} \times \textrm{FWHM}^{-1/2}$ as a function 
of $W_c$ and FWHM in Fig. \ref{fig:fig4}. The contours show that the total SNR of the narrowband survey peaks at 
$W_c$ $\sim$ 965$\pm$2 nm, which indicates that the best central wavelength of the filter bandpass should be around this wavelength range. 
 
\begin{figure}[H]
   \begin{center}
	\includegraphics[width=0.5\textwidth,angle=0]{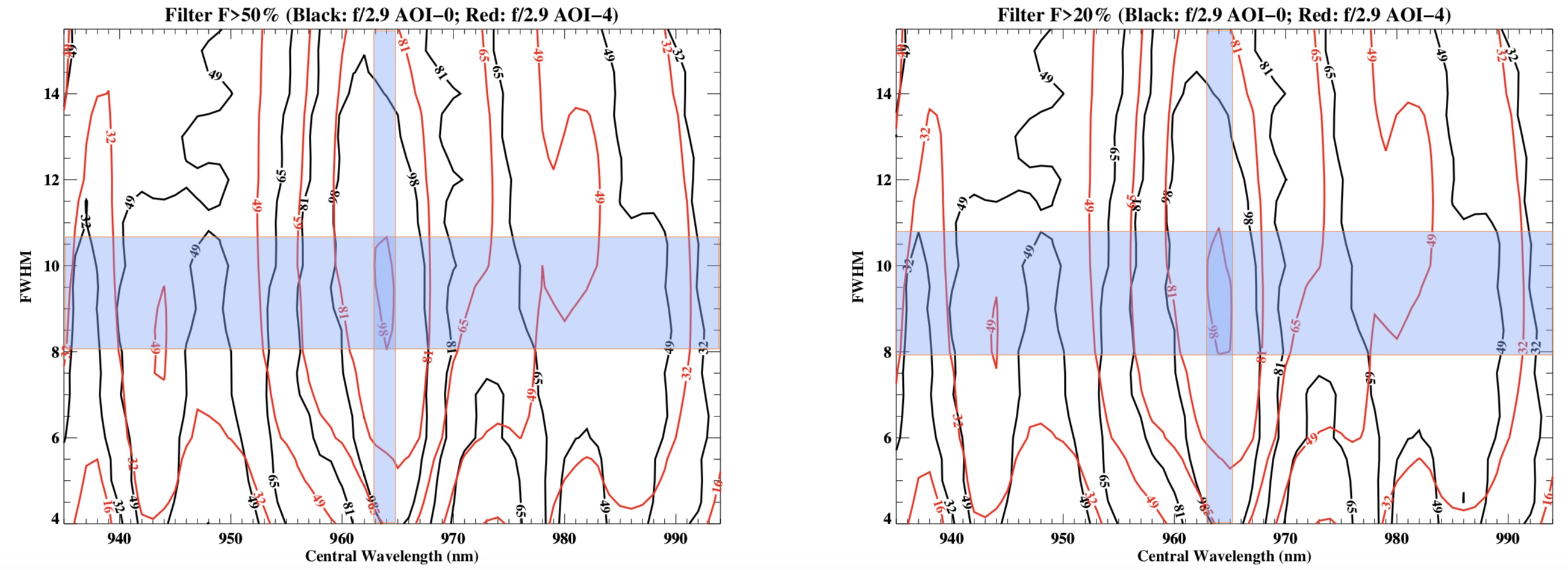} 
 \end{center}
\caption[example]{\label{fig:fig5}Contours of the filter-profile modification parameter $\zeta(\textrm{W}_c, \textrm{FWHM})$ as 
a function of filter parameters $\textrm{W}_c$ and $\textrm{FWHM}$.  
The integration over the filter profile is applied in the wavelength range where the transmission is $\geq$50\% 
(left panel) and $\geq$20\% (right panel) of the peak value, respectively. The black and red contours are based 
on calculations under AOI of 0 degree and 4 degree, respectively. The shaded blue regions roughly mark the 
best parameters of the filter's bandpass.}
\end{figure}

Next we define the filter width FWHM, which is directly connected to the survey volume.  
A wider NB filter design (larger FWHM) would cover a larger survey volume, but the inclusion of more sky background in the 
NB filter would make the survey depth shallower, and hence decreasing the survey efficiency. 
However, when choosing a narrower NB filter, fewer sky background will be covered, but the survey volume becomes smaller. 
Therefore, it is required to balance the survey volume and the survey depth when designing the narrowband filter for DECam. 
The strategy would be the same for future narrowband surveys in NIR regions with dense sky background, 
such as (possible) LSST narrowband surveys for galaxies at $z\gtrsim$ 7.

The survey depth is quantified by the limiting luminosity $L_\textrm{min}$, which is converted from the 5-$\sigma$ 
background RMS signal times the luminosity distance modulus. It is calculated as 
\begin{eqnarray}\label{Eq4}
L_\textrm{min} & = & 4\pi d_L^2 F_{Lim} \\
               & = & 4\pi d_L^2  f_\textrm{Norm} \sqrt{T_\textrm{Exp} \int \Psi_\lambda(W_c,\textrm{FWHM})\textrm{Sky}_\lambda\textrm{QE}_\lambda d\lambda }, \nonumber
\end{eqnarray}
where $d_L$ is the luminosity distance at redshift $z$, and $f_\textrm{Norm}$ is the conversion factor from 
background RMS signal in Eq. \ref{Eq3} to a detection flux limit $F_{\textrm{Lim}}$. Since $f_\textrm{Norm}$ 
is a constant, after integrating over $\lambda$, it is obvious that $L_\textrm{min}$ is a function of  $\textrm{W}_c$ and $\textrm{FWHM}$.

We can quantify the survey efficiency by estimating the number of LAEs which can be detected based on a survey volume and a survey depth. 
Assuming a Schechter luminosity function with characteristic luminosity L$^*$, characteristic density $\Phi^*$ and faint-end slope $\alpha$, 
the estimated number $dN$ of LAEs between redshift $z$ and $z$+$dz$ is caculated as 
\begin{eqnarray}\label{Eq5}
dN(L> L_\textrm{min} & = & \frac{L^0_{min}}{f'_{corr}(\lambda)})  \\\nonumber
 & = & \Phi^* \Gamma(\alpha+1,\frac{L_\textrm{min}}{L^*})\times d\textrm{V}  , \nonumber
\end{eqnarray}
where $L^0_{min}$ is the minimal line luminosity in the case of a boxcar shape, $f^{'}_{corr}(\lambda)$ accounts for the 
effect of a non-boxcar shape, and $\Gamma(\alpha+1,\frac{L_\textrm{min}}{L^*})$ is an incomplete gamma function representing the integration 
over the Schechter luminosity function. 
Because $z + 1$ = $\lambda$/121.567 nm, the survey volume $d$V between redshift $z$ and $z$+$dz$ is 
also a function of wavelength $\lambda$.
In the filter center, $f^{'}_{corr}(\lambda)$ = 1, while $f^{'}_{corr}(\lambda)$ decreases towards the wings of the filter. 
Therefore, the total number of detectable LAEs from a narrowband imaging with filter profile $\Psi_\lambda(W_c,\textrm{FHWM})$ is
an integration over Eq. \ref{Eq5}, which is
\begin{eqnarray}
N(W_c,\textrm{FWHM}) & = & \int_{\Psi_\lambda} \textrm{d}N(L> L_\textrm{min}(\lambda)) \textrm{d}\lambda, \\\nonumber
  & = & N^0(\textrm{W}_c, \textrm{FWHM}) \times  \zeta(\textrm{W}_c, \textrm{FWHM}), 
\end{eqnarray}
where $N^0(\textrm{W}_c, \textrm{FWHM})$ is the number of detectable LAEs with an ideal square boxcar shape filter, and 
$\zeta(\textrm{W}_c, \textrm{FWHM})$ is the modification on that number when the real NB filter profile is not in that shape. 
By adjusting the exposure time $T_\textrm{Exp}$, we can reach $L_\textrm{Lim}$ = 0.5 $L^*$, a typical survey depth, then assuming the faint-end slope
$\alpha$ = -2\footnote{Through an ultra-deep spectroscopic measurement of the faint-end slope for LAEs at $z$ = 5.7, 
\citet{Dressler+2015} constrained the faint-end slope to -2.35 $< \alpha <$ -1.95 (1$\sigma$). So we choose $\alpha$ = -2 in the calculation.} and integrating over the luminosity range of 0.5L* to 10L* with the gamma function, the relative number $\zeta(\textrm{W}_c, \textrm{FWHM})$ of observed LAEs under different parameters of the narrowband filter is derived. 

In Fig. \ref{fig:fig5} we plot the contour of the relative number $\zeta(\textrm{W}_c, \textrm{FWHM})$ as a function of filter parameters $\textrm{W}_c$ and $\textrm{FWHM}$. 
The integration over the filter profile is applied in the wavelength range where the transmission is $\geq$ 50\% and $\geq$ 20\% of the peak value, respectively. 
The contours of the relative number $\zeta(\textrm{W}_c, \textrm{FWHM})$ show peak values  
at central wavelength of 964$\pm$1 nm and FWHM value of 9$\pm$1 nm. It indicates that with these filter parameters 
 the survey efficiency can be maximized.

\section{Manufacture of the DECam Narrowband Filter NB964}\label{sec:made}

\subsection{Recent Technological Advances}\label{sec:made:tech}

Thanks to the rapid development in coating technologies, narrow bandpass filters with superior uniformity over a large area have become practical.
This provides opportunities for narrowband imaging surveys with large field of views such as large focal planes of DECam and HSC. 
Optical companies like Materion have built new facilities to manufacture this kind of filters \citep{Mooney+2014}. 
Among different deposition processes, Materion chooses the reactive magnetron sputter deposition because of its low risk, high stability, 
high precision, and strict uniformity \citep[for details, see chapter 4 of][]{Mooney+2014}.  
In this process, material is sputtered from a cathode in a new coating chamber specially designed for large optics at Materion.  
This sputtered material bombards the substrate to be coated, where it is deposited.  
It provides sufficiently accurate control on the thickness of coating layers
that bandpass features can be controlled to $0.25\%$ accuracy or even better for filters with diameters of 600mm.

\subsection{Bandpass Specification}\label{sec:made:bandpass}
We submitted the basic parameters of the NB964 filter as well as technical requirements to Materion in 2014. 
The technical specifications are summarized as below.

\subsubsection{General Specifications of NB964}\label{sec:made:bandpass:1}
\begin{description}
\item[Substrate dimensions] 620$^{+1.0}_{-0.0}$ mm diameter circle.
\item[Substrate material] Fused silica.
\item[Minimum Clear Aperture] 600mm diameter circle, concentric with substrate diameter.
\item[Thickness] Total 13.0$^{+1.0}_{-0.0}$ mm, inclusive of substrates and coatings.
\item[Surface parallelism] 30 arc-seconds or better. 
\item[Flatness] Transmitted wave front error $<\lambda$/4 in 125mm diameter sub-aperture.
\item[Bubbles and inclusions in substrate] Bubble class zero with a requirement of $< 0.1\hbox{mm}^2$ per 100 cm$^3$. 
\item[Striae] Grade A with a requirement of $<$ 30nm. 
\item[Pinhole limitations]
\begin{description}
\item 
\item [$>60 \mu m $ diameter]  None allowed;
\item [$30$ to $60 \mu m$ diameter]  $\leq$1 per 50mm diameter region of the filter;
\item [$<30 \mu m$ diameter] In each 10mm diameter within the clear aperture, the area of the total pinholes 
 must be less than the equivalent area of a $30\mu m$ pinhole.
\end{description}
\item[Surface finish] 60-40 Scratch-Dig.
\item[Durability] Mil-C-48479 (moderate abrasion).
\item[AR coatings and edge sealing]
 Both surfaces must be anti-reflection coated
  with as hard a coating as possible. The filter edges should
  be hermetically sealed against moisture penetration.
Also, filter edges should be made as black as possible to help mitigate 
the scattered light from the filter edges.
\item[Radioisotope limits in filter substrate] 
U $<$ 0.8 ppm, Th $<$ 2.5 ppm, and K $<$ 0.03\% (by weight).
\item[Radioisotope limits in filter coatings]
U $<$ 80 ppm, Th $<$ 250 ppm, and K $<$ 3\% (by weight).
\item[Index of refraction of coatings] 
We expect to minimize the bandpass shift
with the increasing angle of incidence.  Since the transmitted wavelength
goes as $\sqrt{1-(sin\theta/n)^2}$, 
the effective $n$ should be as high as possible.
\item[Edge masking] 
 The out-of-band blocking should apply equally to 
the part outside of the clear aperture.
\item[Environmental conditions]
\begin{description}
\item 
\item [\it Operating range] Temperature $-7$ to $+27^\circ$C; 
humidity $0$ -- $60\%$ (dry nitrogen environment); elevation 2200m (7220 ft), 
pressure 570 -- 590 torr.
\item [\it Survival range] Temperature -56$^\circ$C to $+50^\circ$C; 
humidity 0 -- 100\%; elevation 4500m (15000 ft),
pressure 410 -- 780 torr.
\end{description}
\item[Operating focal ratio] f/2.9.
\item[Beam angles of incidence] 0 to 4$^\circ$ for f/2.9 beams
(maximum range 0--12$^\circ$).
\end{description}

\subsubsection{Passband Specifications}\label{sec:made:bandpass:2}

\begin{figure*}[t]
   \begin{center}
   \includegraphics[width=\textwidth,angle=0]{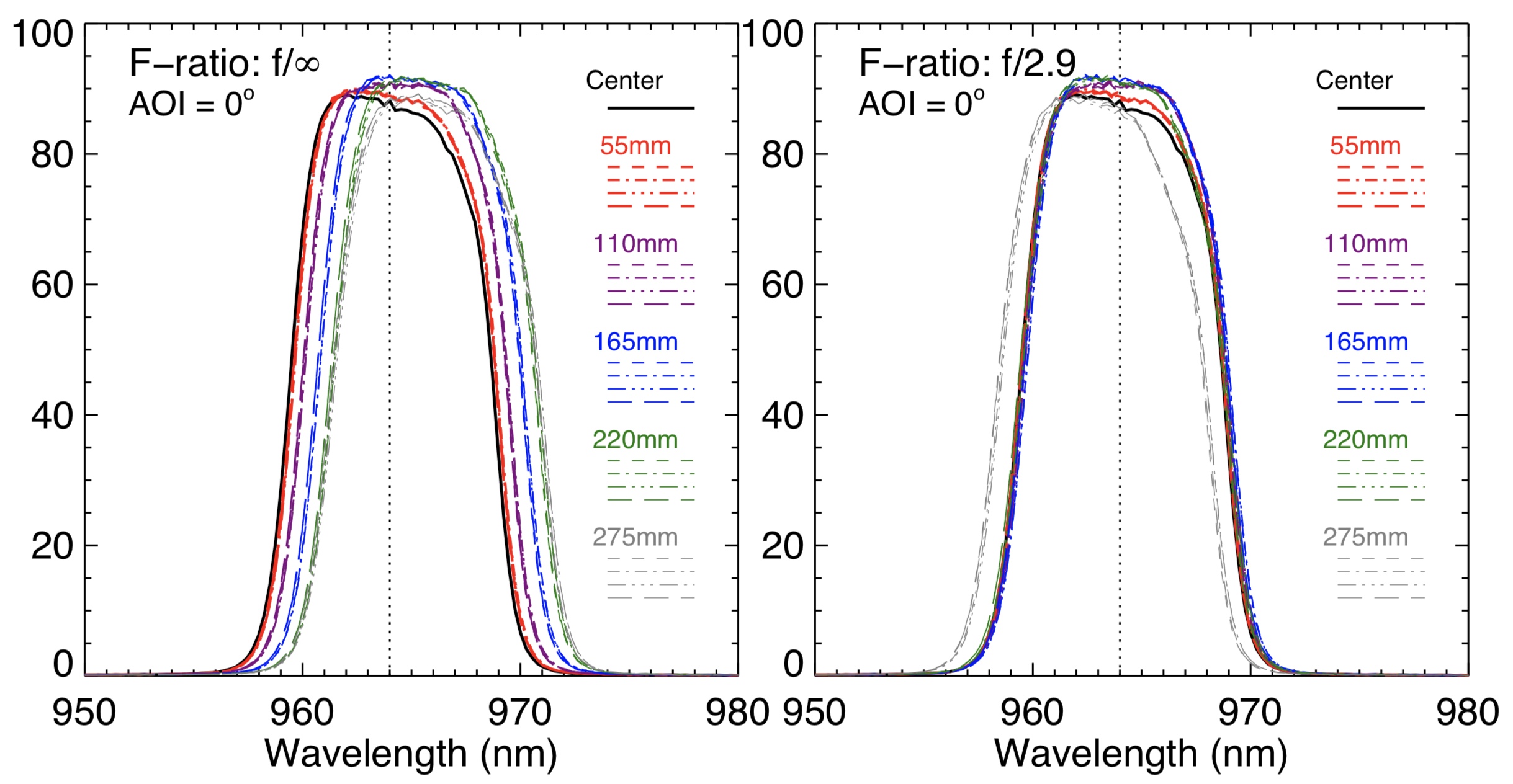}
   \end{center}
   \caption[example] 
{ \label{fig:fig6} 
Transmission curves of the NB964 filter measured at the normal incidence with F-ratio f/$\infty$ \& AOI $\theta$ = 0$^\circ$ (left panel) and derived for F-ratio f/2.9 \& AOI $\theta$ = 0$^\circ$ (right panel).
The measurements were performed at Materion with a spectrometer at AOI $\theta$ = 0$^\circ$, F-ratio f/$\infty$,  and a room temperature of 25 $^\circ$C. 
The measurements were taken at the filter center (black solid lines) and various off-center positions along 4 different directions (shown as dashed, two types of dash-dotted, and long-dashed lines) from the center, 
at five radii with 55-mm increment marked as red, purple, blue, green and grey colors, respectively. The measurements with F-ratio f/2.9 on the right panel were derived based on Eq. \ref{eq2}.}
\end{figure*} 

\begin{itemize}
\item The passband is specified for f/2.9 beam at normal incidence, at operating temperature $\sim5^\circ$ C.  
\item Central wavelength:  $\lambda_c = 964 \pm 2.0 \nm$
 to apply at all locations within the clear aperture (for normal-incidence,
 f/2.9 beam).
\item Average FWHM:  $9.0^{+ 1.0}_{-1.0} \nm$ (for normal-incidence,
 f/2.9 beam).
\item The minimum allowable peak transmission ($\tmax$) of the filter is 80\%.
 This is to apply for all spatial positions within the clear aperture.
\item Spatial variations in $\tmax$ are to be less than 5.0\% of $\tmax$
 for all positions in the clear aperture.
\item The transmission within the wavelength range at $\gtrsim$90\% of the peak transmission (BW90) should not fall below 85\% of the peak
 transmission.
\item The bandwidth at 10\% of the peak transmission (BW10) should be
 no larger than 150\% of the FWHM for a normal-incidence, f/2.9 beam.
\item Ghost image amplitude:  To quantify the brightness of  
ghost images caused by double reflections within the filter,
we define the ``fractional ghost flux'' as 
$$\hbox{FGF} = {\int_0^\infty T_1 R_2 R_1 T_2 d\lambda
\over
\int_0^\infty T_1 T_2 d\lambda},
$$
where $T_1$ and $R_1$ are the transmissivity and reflectivity of the front side of the filter as a function of wavelength, respectively, 
while $T_2$ and $R_2$ are for the back side of the filter. We require $\hbox{FGF} < 0.01$.
\item Blocking outside of passband:\\
\begin{tabular}{lll}
300 to 1050 nm: & $<10^{-4}$ & (0.01\%) \\
1050 to 1200 nm: & $<10^{-3} $ & (0.1\%) 
\end{tabular}
\end{itemize}

\section{Tests of the DECam Narrowband Filter NB964}\label{sec:test}

Materion finished the manufacture of the narrowband filter NB964 in the summer of 2015. Based on a series of tests, 
we find that the technical parameters and specifications of NB964 agree well with the design introduced in sections \S\ref{sec:method} and \S\ref{sec:made}. 
These tests include the lab tests at Materion, the on-site 
tests with DECam at CTIO, and on-sky observational tests, which are introduced as below.

\subsection{Lab Tests at Materion}\label{sec:test:MaterionTest}

Materion obtained the lab test results of NB964 before the filter was ready to ship to CTIO.
The lab tests of NB964 include measurements of the transmission curves, the blocking curves, and the basic filter parameters. 
The transmission curves at various positions of NB964 were measured by a spectrometer with a light beam at the normal incidence and room temperature of 25 $^\circ$C, which are 
plotted in the left panel of Fig. \ref{fig:fig6}. 
The peak transmissions $T_\textrm{max}$ at different positions are $\sim$90\% on average and show spatial deviation (the maximum difference) of $3.83$\%, 
which meet the design requirements ($T_\textrm{max}$ $>$ 80\% and spatial variation $<$ 5\%) listed in Sec. \ref{sec:made:bandpass:2}. 
With AOI $\theta$ = 0$^\circ$ and F-ratio f/$\infty$, the transmission curves at different locations of the coated filter NB964 show systematic red-shifts 
in wavelength as a function of increasing radii. The central wavelength $W_c$ shows slight shifts toward the red side from 964nm to 966nm when moving radially outward from the center. 
This red-shifts adjustment is made for the purpose of observing with the actual focal ratio, as discussed in the following.  

Based on Eq. \ref{eq2}, we derive the transmission curves of NB964 with AOI $\theta$ = 0$^\circ$ and F-ratio f/2.9, which is plotted on the right panel of Fig. \ref{fig:fig6}. 
This plot shows that the transmission curves with f/2.9 measured at radii of $R  =$ [55, 110, 165, 220] mm 
in the 4 different directions all match well with each other with wavelength shifts $\lesssim$0.3 nm.  
Although the transmission curves measured at the radius of $R = 275$ mm show a systematic blue-shift by $\sim$1 nm, it still meets our requirement of 964$\pm$2 nm for $W_c$. 
The blue-shift at larger radii is caused by the inadequate red-shift under the condition f/$\infty$ (grey lines on the left panel of Fig. \ref{fig:fig6}). 
The red-shifts with radii represent the special treatment for the operational focal ratio f/2.9, so that 
the transmission profile of NB964 when observing with DECam should be consistent at different locations of the filter. 

Since the operating temperature of the NB964 filter is 5 $^\circ$C, Materion provided the calculated filter parameters under the condition of 5 $^\circ$C. 
The resulting central wavelength $W_c$ and FWHM are 964.2 nm and 9.2 nm, respectively.

\subsection{On-site Tests with DECam}\label{sec:test:DECamTest}

The filter was shipped to CTIO and installed in the DECam filter box in late 2015. The observation with this filter started in December 2015. 
Later on Oct. 24 2016 NOAO staff  A.K. Vivas used DECal\footnote{DECal website: http://instrumentation.tamu.edu/DeCal.html}\citep{Marshall+2013}, 
 a spectrophotometric calibration instrument implemented in the Dark Energy Survey (DES) project, to measure the 
 on-site transmission of NB964 with DECam. DECal uses a 2nm wide tunable source to illuminate the dome flat which allows 
 us to measure the transmission curve of NB964 as a function of wavelength. 
The 2 dimensional focal plane responses at different wavelength bins are plotted in Fig. {\ref{fig:fig7}.

    \begin{figure}[t]
   \begin{center}
   \includegraphics[width=0.5\textwidth,angle=0]{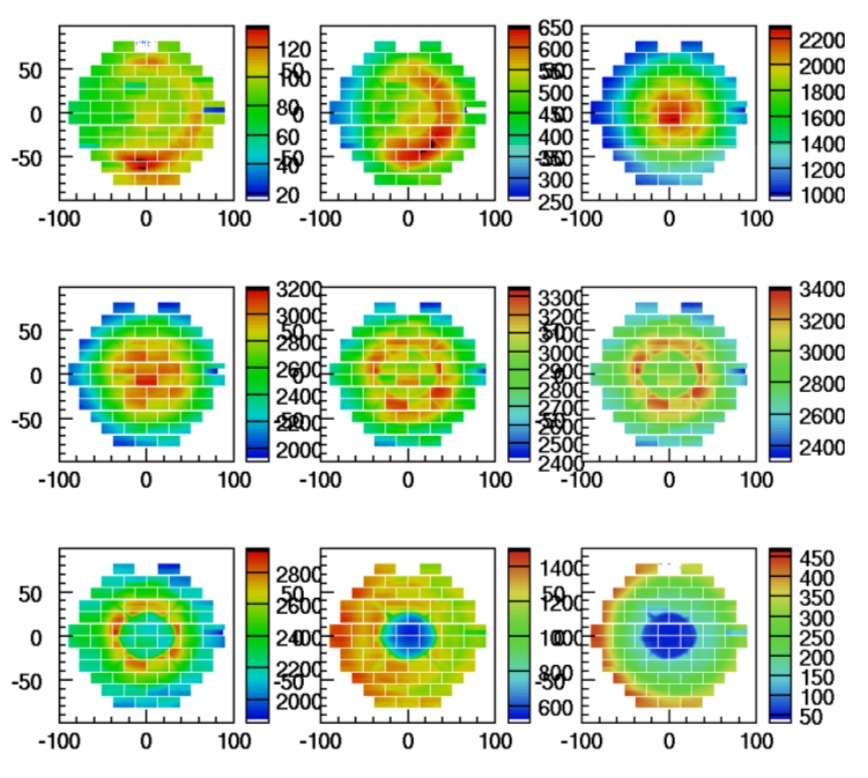}
   \end{center}
   \caption[example] 
{ \label{fig:fig7} 
Focal plane responses measured by DEcal photodiode (ON - OFF intensity normalized for CCD gains)
at 956, 958, 960 nm (top row, from left to right), 962, 964, 966 nm (middle row), 968, 970, 972 nm (bottom row), respectively. }
\end{figure}

\begin{figure}[t!]
   \begin{center}
   \includegraphics[width=0.5\textwidth,angle=0]{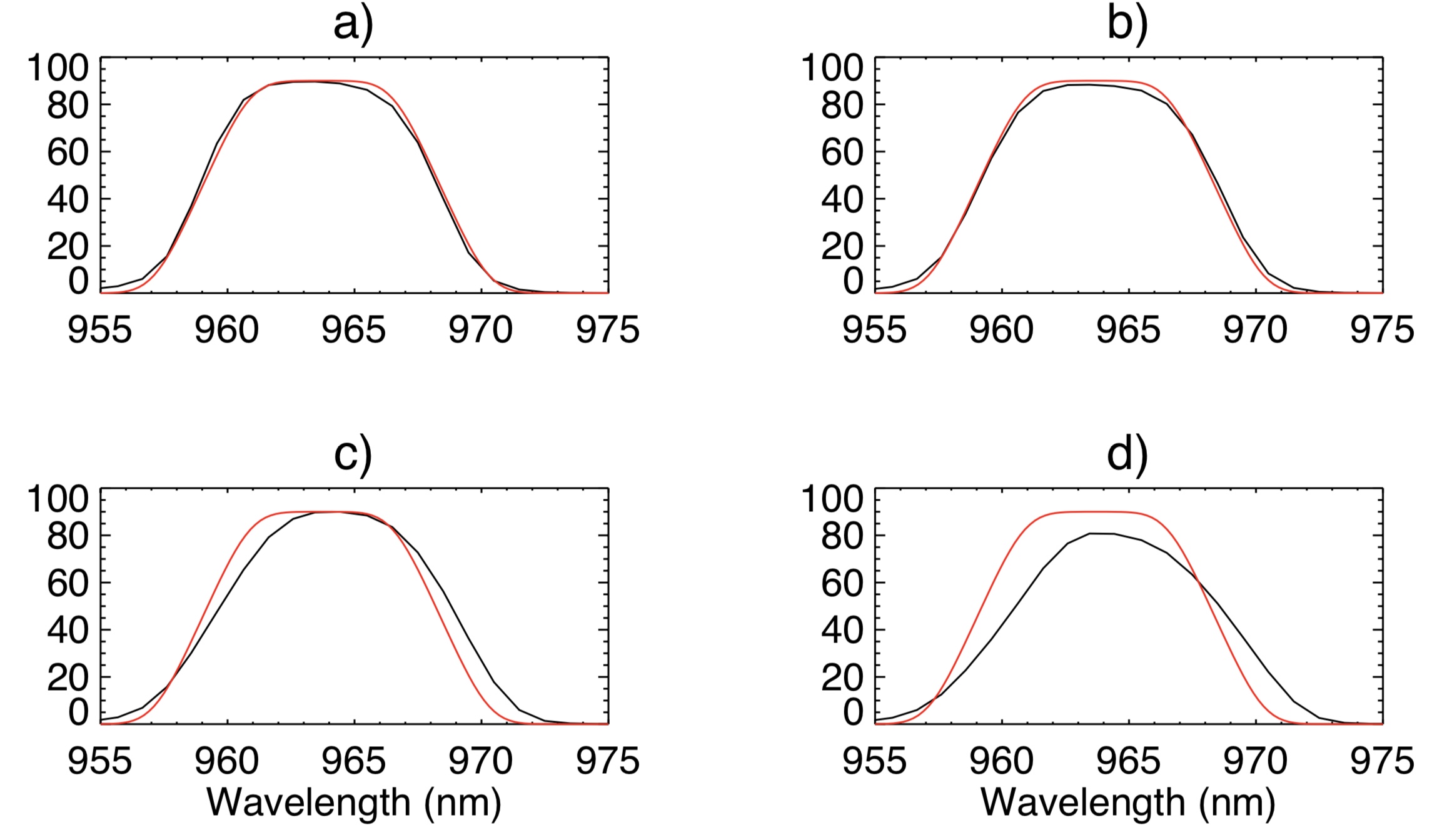}
   \end{center}
  \caption[example] 
{ \label{fig:fig8} 
  Throughput measured on-site by DECal averaged in four DECam annuli (a: innermost 2 CCDs, 
b:  annulus next to the innermost, c: annulus next to the outermost, and d: outermost annulus) 
of the focal plane shown in Fig. \ref{fig:fig7}. 
The red curves show the modeled NB964 profile with f/2.9 and AOI=0$^\circ$. 
}
\end{figure}

  \begin{figure}[t]
   \begin{center}
   \includegraphics[width=0.5\textwidth,angle=0]{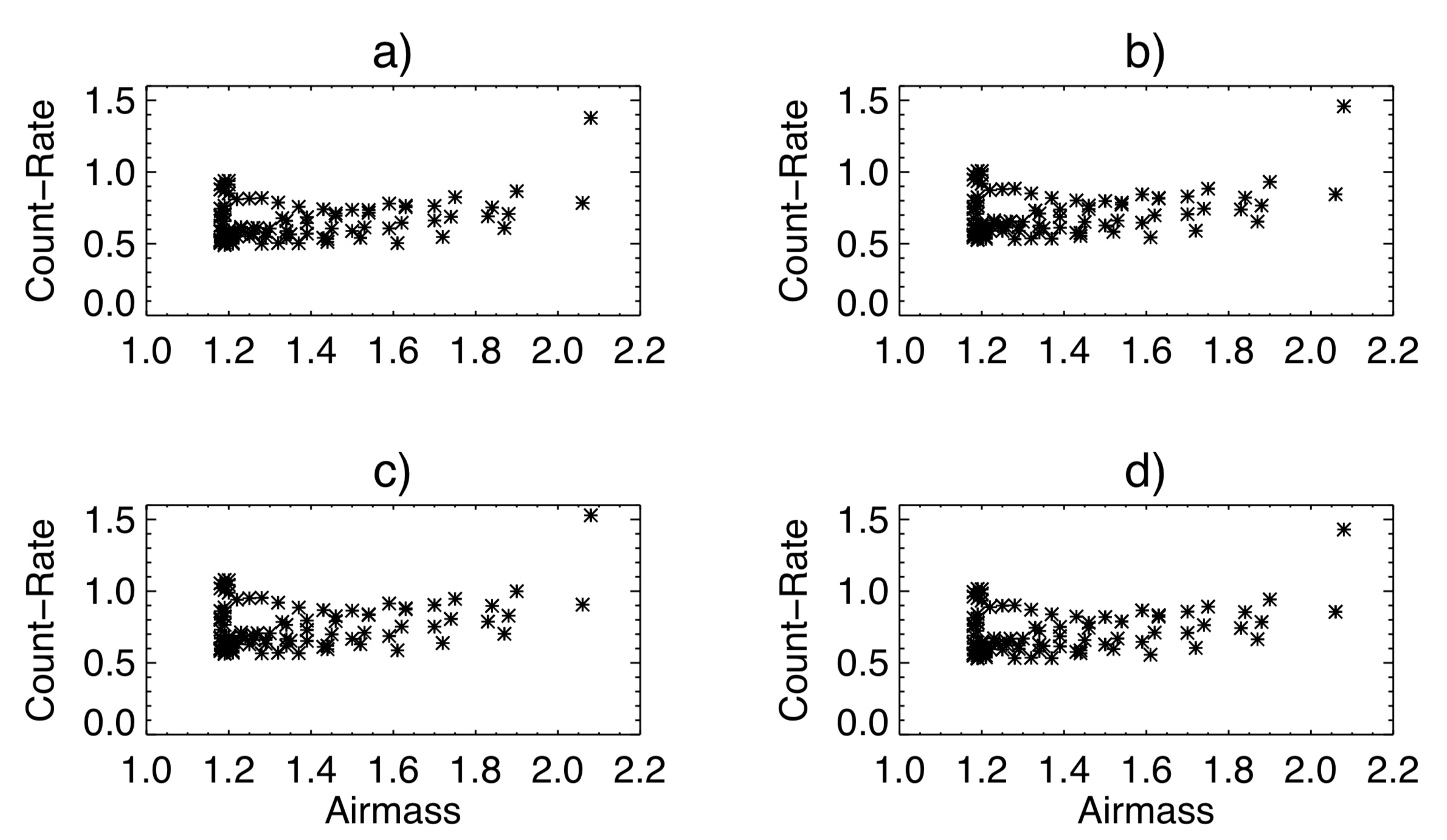}
   \end{center}
  \caption[ ] 
{ \label{fig:fig9} 
Sky background count-rates [unit: ADU/s] of NB964 as a function of the airmass. 
The sky background count-rates are calculated from 98 NB964 individual exposures averaged in four annuli (a: innermost 2 CCDs, 
b:  annulus next to the innermost, c: annulus next to the outermost, and d: outermost annulus) of the focal plane, which have the same definition as in Fig. \ref{fig:fig8}.}
\end{figure} 

  \begin{figure}[th]
   \begin{center}
   \includegraphics[width=0.5\textwidth,angle=0]{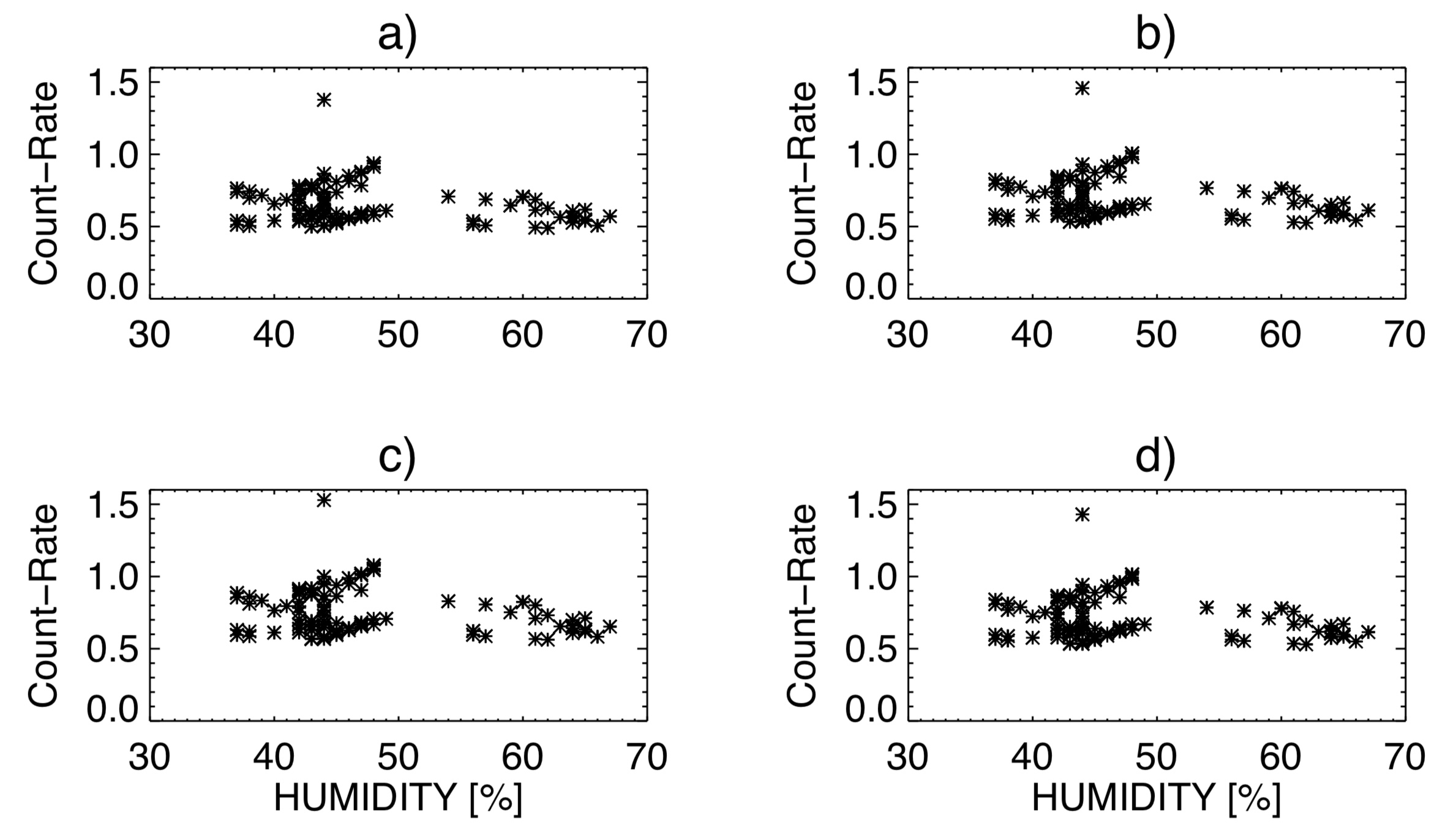}
   \end{center}
   \caption[ ] 
{ \label{fig:fig10} 
Sky background count-rates [unit: ADU/s] of NB964 as a function of the humidity. 
The sky background count-rates are calculated in four annuli (same definition as in Fig. \ref{fig:fig8}) of the focal plane. }
\end{figure}

In Fig. \ref{fig:fig8} we show the transmission curves averaged in different annuli (innermost 2 CCDs, 
annulus next to the innermost, annulus next to the outermost, and the outermost annulus) of the focal plane.
For comparison, we also plot the calculated NB964 filter profile with f/2.9 and AOI=0$^\circ$. Obviously, 
in the inner two annuli the transmission curve (Fig \ref{fig:fig8}-a and \ref{fig:fig8}-b) 
each matches very well with the modeled filter profile and the Materion test (Fig. \ref{fig:fig6}), while in 
the outer two annuli the throughputs shift red-ward slightly (by $\sim$0.7 nm) and in the meantime become wider. 
The on-site tests are made through the complete telescope optical system, and we suspect that the difference 
between that and the measurements made at Materion is due to the properties of the DECam optical 
corrector \citep{Doel+2008}. It could also be due to the non-uniform dome flat on the outside of the focal plane, 
where AOI $\neq$ 0$^\circ$. Therefore we should use a sky/star flat in the data reduction pipeline. 
Currently we use the z-band star flat.

\subsection{On-sky Observations with DECam}\label{sec:test:ObsTest}
The LAGER survey with NB964 on DECam started in Dec. 2015. It is designed to select more than 
a hundred $z\sim$7 LAEs over an area of 12 deg$^2$ in four fields (total Volume: 8 $\times 10^6$ cMpc$^3$). 
Currently, LAGER has collected 123 hrs NB964 narrowband imaging in four fields 
(CDF-S, COSMOS, WIDE-12 and GAMA-15) taken during 18 nights in 2015-2018, making it the 
largest narrowband survey for LAEs at $z\gtrsim$7 so far. The deepest NB964 imaging is done in COSMOS, 
where we have obtained 47.25 hrs NB964 exposure in a 3 deg$^2$ field. The exposure time per NB964 frame is 900s. 
Consecutive exposures are dithered by $\sim$100\arcsec\ so that chip gaps and bad pixels do not 
lead to blank areas in the final stacks. We also take 0.5-1 hr z-band and $Y$-band exposures per field to exclude possible transients. 
The NB964, DECam z-band and DECam $Y$-band data were scientifically reduced by DECam Community Pipeline \citep{Valdes+2014}.
Here we summarize the observational characteristics of the NB964 filter taken in Feb. 4-9, 2016 .

\subsubsection{On-sky Count-rates of NB964}
The count-rate values of the sky background in individual frames of NB964 are in the level of 0.5-1.0 ADU/s. 
The typical GAIN value of DECam is 4 e$^-$/ADU. Therefore, the sky values are 1800-3600 e$^-$ in one 
NB964 exposure of 900s. These values correspond to the NB964 sky brightness of 19.8-19.1 mag, which 
is typically 1 mag fainter than the sky brightness in z-band with 50 deg to the bright moon (18.2-18.7 mag, 
extracted from DECam ETC\footnote{DECam ETC website: http://www.ctio.noao.edu/noao/content/Exposure-Time-Calculator-ETC-0
}). As noted in the DECam handbook, some amplifiers are non-linear below $\sim$1000 e$^-$. 
While this is corrected in the reduction pipeline, we prefer to operate out of this regime by taking long exposures 
($\geq$ 900s) with N964 so that there is no chance of photometry of very faint objects being compromised.

In Table \ref{ctrcomp} we list simulated count-rate ratios of the sky background between NB964, DECam z-band and 
DECam Y-band under different sky conditions.  The sky background count-rate of NB964 $\frac{CR(NB964)}{CR_0(NB964)}$, 
which is normalized by the value with airmass 1.0 and water vapor 2.3mm, appears to be weakly correlated both 
with the airmass and with the water vapor. However, we do not see clear correlation signals in the observed data 
either with the airmass (Fig. \ref{fig:fig9}) or with the humidity ( Fig. \ref{fig:fig10}).

\begin{table}[htp]
\caption{Simulated count-rate (CR) ratios of the sky background between NB964, DECam z-band, and DECam Y-band under different sky conditions. }
\begin{center}
\begin{tabular}{|cc|ccc|}\hline\hline
Water  & Airmass & $\frac{\textrm{CR(NB964})}{\textrm{CR}_0(\textrm{NB964})}$ & $\frac{\textrm{CR(NB964)}}{\textrm{CR(z)}}$ & $\frac{\textrm{CR(NB964)}}{\textrm{CR(Y)}}$ \\\hline\hline
2.3 mm & 1.0 & 1.000  & 0.0387 & 0.0617 \\
2.3 mm & 1.5 & 1.086  &  0.0371  & 0.0511 \\
2.3 mm & 2.0 & 1.171 & 0.0359 & 0.0446 \\\hline
10 mm & 1.0 & 0.870 & 0.0363  & 0.0560 \\
10 mm & 1.5 & 0.913 & 0.0344 & 0.0451\\
10 mm & 2.0 & 0.955 & 0.0330 & 0.0383 \\\hline
\end{tabular}
\end{center}
\label{ctrcomp}
\end{table}

\begin{figure}[t]
   \begin{center}
   \includegraphics[width=0.5\textwidth,angle=0]{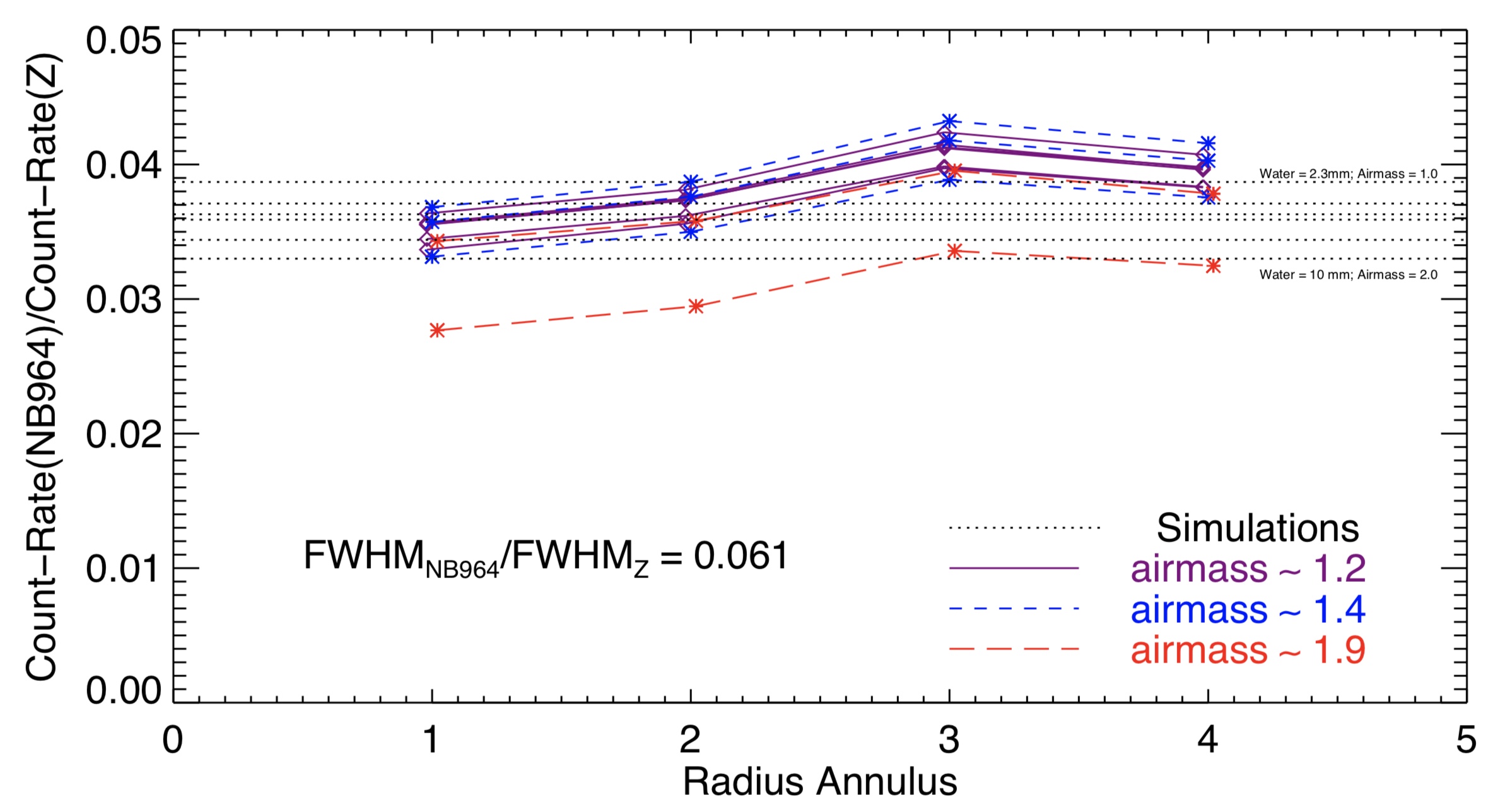}
   \end{center} 
   \caption[example] 
{ \label{fig:fig11} 
 Observed count-rate ratios of the sky background in NB964 and DECam z-band as a function of radial annulus (from inside to outside, same definition as in Fig. \ref{fig:fig8}). 
The simulated count-rate ratios with different values of water and airmass listed in Table \ref{ctrcomp} are plotted as dotted lines. 
}
\end{figure}

\subsubsection{Comparison of the On-sky Performance between NB964 and z-band}

 We choose 11 pairs of exposures when NB964 and z-band observations were taken consecutively, and plot
 the ratio of the sky count-rate in each pair in Fig. \ref{fig:fig11}. While the ratio of the FWHM of NB964 to that 
 of the z-band is 6.1\%, the sky photons collected by the 
 NB964 filter to that by the z-band filter is only $3.8\pm0.3$\%. 
  It is in a good agreement with the designed values, which are listed in Table \ref{ctrcomp}.

 The variation on the ratio of the observed count-rate of NB964 to that of the z-band can be explained by the 
 joint effect of water-vapor and airmass, as seen from Fig. \ref{fig:fig11} and Table \ref{ctrcomp}. However, 
 the scattered light from the moon and thin clouds would also affect the background radiation, either by increasing 
 the background continuum or reducing the amount of OH emission to the telescope. Their effects need to be 
 analyzed with good weather records at CTIO, which is difficult to get currently.

\subsubsection{Filter Profile Explored with the Low-Redshift Spectroscopic Sample}

As an independent and first-order check, we explore the NB964 filter profile with the spectroscopic sample 
of NB964 selected low-redshift emission-line galaxies, such as H$\alpha$ and [O\,\textsc{iii}] emitters. These 
spectroscopically confirmed low-redshift emission line galaxies are firstly selected with $z$ - NB964 $\geq$ 0.3 \& NB964 $<$ 25, 
and then matched with the G10/COSMOS redshift catalogue \citep{Davies+2015, Andrews+2017}. 
About 210 H$\alpha$ emitters and 40 [O\,\textsc{iii}] emitters selected from LAGER have matched counterparts 
with high resolution spectroscopic redshifts or reliable PRIMUS low resolution spectroscopic redshifts in the G10/COSMOS redshift catalogue. 

In Fig. \ref{fig:fig12} we show the redshift distributions of LAGER H$\alpha$ and [O\,\textsc{iii}] emitters, which match well with the scaled NB964 
filter profiles. It is a nice consistency check on the filter profile. However, the size of the current spectroscopic sample is not large enough to 
give a more detailed information, e.g.,, the spatial variation of the filter profile. Future large spectroscopy surveys such as Subaru/PFS 
\citep[Prime Focus Spectrograph,][]{Sugai+2012} and VLT/MOONS \citep[Multi Object Optical and Near-infrared Spectrograph,][]{Cirasuolo+2014} will 
enlarge the spectroscopic sample, which could allow us to check the spatial variation of the filter profile.

\begin{figure}[!t]
   \begin{center}
   \includegraphics[width=0.45\textwidth,angle=0]{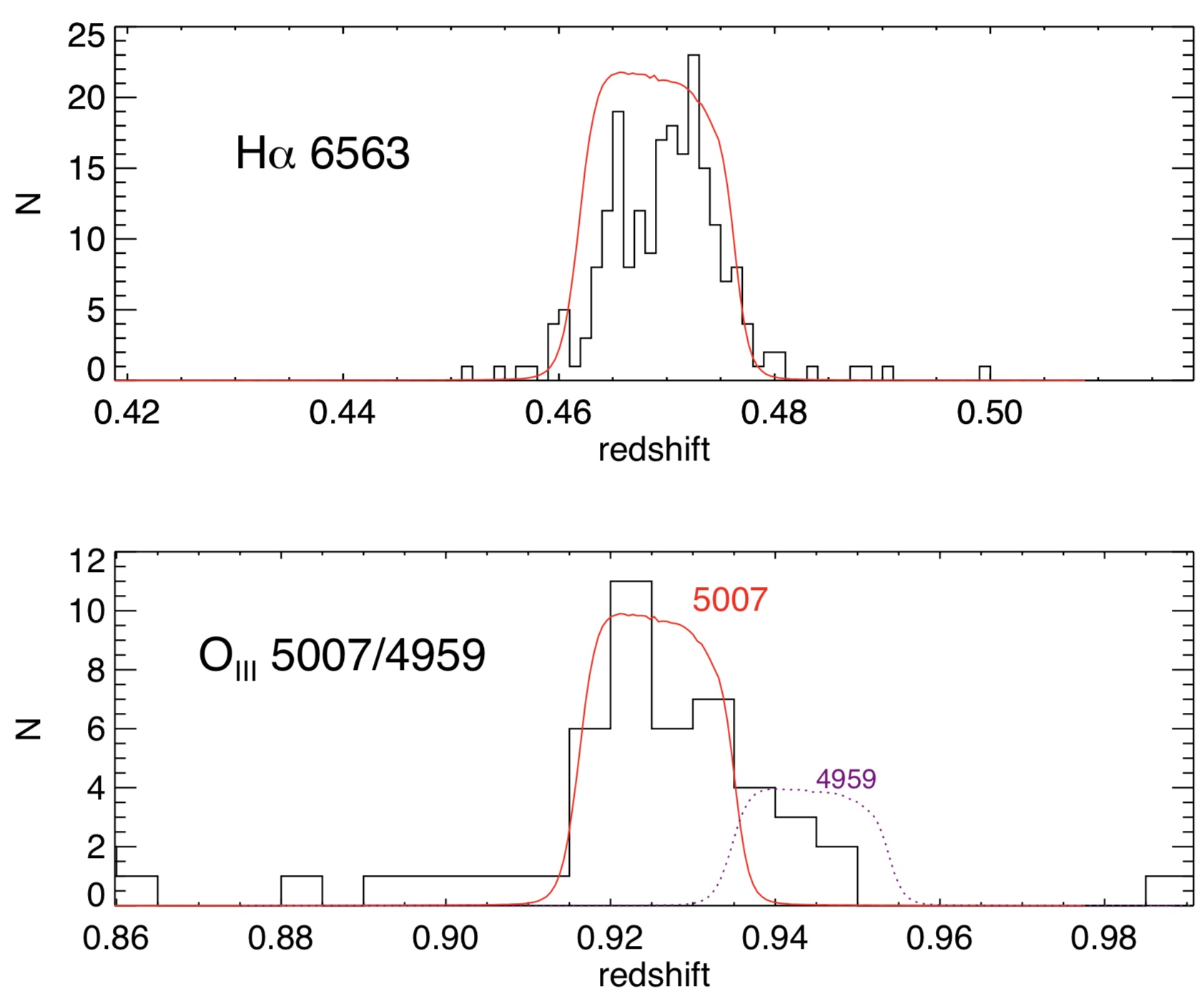}
   \end{center}
   \caption[example] 
{ \label{fig:fig12} 
Redshift distributions of the LAGER H$\alpha$ emitters (top panel) and [O\,\textsc{iii}] emitters (bottom panel) 
compared to the scaled NB964 filter profiles at the corresponding redshifts (red and purple curves). 
There are about 210 H$\alpha$ emitters and 40 [O\,\textsc{iii}] emitters selected from LAGER (criteria: $z$ - NB964 
$\geq$ 0.3 \& NB964 $<$ 25) and matched with the G10/COSMOS redshift catalogue \citep{Davies+2015, Andrews+2017}.  }
\end{figure}

\subsection{Discussion on the Impact of the Filter Profile}\label{sec:discussion}

As introduced in Section \ref{sec:method}, the profile of the narrowband filter is crucial to the efficiency of the 
ground-based narrowband imaging survey for LAEs in the EoR. Compared with previous narrowband filters 
with a close central wavelength, NB964 filter has an excellent band shape more like a square "boxcar" rather than a gaussian. 
Narrowband filters with the gaussian shape would systematically underestimate the \lya\ fluxes when the 
observed \lya\ wavelength is located in the filter wings. Although sometimes this effect can be statistically 
corrected for the sample as a whole, it is not the case for rare objects that only fall into the wing area. 
For example, the 3 most luminous LAEs in the COSMOS field reported by \citet{Zheng+2017} with NB964 were 
also detected by \citet{Itoh+2018} with their narrowband filter NB973 on Subaru/HSC, but the images are much fainter in NB973 than in NB964. 
Spectroscopic confirmations of these 3 LAEs \citep{Hu+2017} indicate that their \lya\ lines fall into the wings of NB973 which has a gaussian-like shape. 
Therefore, the actual \lya\ luminosity of the 3 bright LAEs can not be obtained  from the NB973 image, if without the spectroscopic redshifts.  
In contrast, such rare objects would be more easily caught with the filter having a square boxcar-like shape.
That's why the bright-end excess of LAEs at $z\sim$ 7 in the COSMOS field reported by \citet{Zheng+2017} with NB964 was not revealed by \citet{Itoh+2018} with NB973. 

Another important issue on the filter profile is about the wavelength-shifts at different locations of the filter.
If there exist wavelength shifts, nearby sky emission lines can be included so that the survey depth is dependent on the filter positions. 
In the inner regions of the filter, our tests in Section \ref{sec:test:DECamTest}
confirm a consistent transmission profile when observing with DECam, while in 
the outer regions the transmission profiles shift by $\sim$0.7nm compared to that in the filter center. 
The area-weighted mean profile of NB964 shows a red-shift of 0.6 nm when compared to the profile in the center. 
This is similar to the HSC narrowband filters NB921, NB926, NB973 and NB1010, which have wavelength-shifts of 
about 1.0, 1.0, 1.5 and -0.3 respectively, when comparing the area-weighted mean profile with that in the filter 
center\footnote{https://www.subarutelescope.org/Observing/Instruments/HSC/sensitivity.html}.

\section{Conclusion}\label{sec:conclusion}

Large filters with strict uniformity are now technologically feasible.  
Here we explore in detail the design and tests of a
narrow band filter NB964 that optimizes the survey efficiency for $z\sim$ 7 \lya\ galaxies with DECam. To that end, we 
have taken into account the sky absorption and emission background, the DECam CCD QE, as well as the galaxy 
number density models. The optimized narrowband filter NB964 has a central wavelength of 964.2 nm and a width of 9.2 nm with 
f/2.9 and 0$^\circ$  AOI. The parameters of the filter have been tested by several methods, which match well with its design. 

With the NB964 filter, a DECam narrowband imaging survey named LAGER (Lyman alpha galaxies in the epoch of reionization) 
has been ongoing. LAGER has helped to discover 22 new $z\sim 7$ LAEs in 2017 \citep{Zheng+2017}. 
A larger sample of $z\sim$7 LAEs will be released soon \citep{Hu+2019}. 
The excellent performances of the NB964 filter have guaranteed the high efficiency and reliability of LAGER in 
detecting galaxies in the epoch of reionization. These large numbers of newly detected LAEs are very helpful for us to reveal the nature of patchy reionization at $z \sim$ 7.


\section*{Acknowledgments\label{sec:acknowledgements}}

We acknowledge financial support from National Science Foundation of China (grants No. 11421303) 
and National Program for Support of Top-notch Young Professionals for covering the cost of the NB964 narrowband filter. 
Z.Y.Z. is sponsored by Shanghai Pujiang Program, the National Science Foundation of China (11773051), and the 
China-Chile Joint Research Fund (CCJRF No. 1503). Work on this project by JER and SM has been supported in 
part by grant AST-1518057 from the United States National Science Foundation. J.X.W. thanks support from National 
Basic Research Program of China (973 program, grant No. 2015CB857005), National Science Foundation of China (NSFC 11890693), and CAS Frontier 
Science Key Research Program QYCDJ-SSW-SLH006. F. B. thanks support from CONICYT Project BASAL AFB-170002.

We thank Materion company for the manufacture of the NB964 filter, which makes the {\it LAGER} project possible. 
We greatly appreciate the kind support from staffs at NOAO/CTIO to make our observations successful. 

Based on observations at Cerro Tololo Inter-American Observatory, National Optical Astronomy Observatory (NOAO 
PID: 016A-0386, PI: Malhotra, and CNTAC PIDs: 2015B-0603 and 2016A-0610, PI: Infante), which is operated by the 
Association of Universities for Research in Astronomy (AURA) under a cooperative agreement with the National Science Foundation. 
Based in part on data collected at the Subaru Telescope and obtained from the Subaru-Mitaka-Okayama-Kiso Archive System (SMOKA), 
which is operated by the Astronomy Data Center, National Astronomical Observatory of Japan.

This project used data obtained with the Dark Energy Camera (DECam), which was constructed by the Dark Energy Survey 
(DES) collaboration. Funding for the DES Projects has been provided by the DOE and NSF (USA), MISE (Spain), STFC (UK), 
HEFCE (UK). NCSA (UIUC), KICP (U. Chicago), CCAPP (Ohio State), MIFPA (Texas A\&M), CNPQ, FAPERJ, FINEP (Brazil), 
MINECO (Spain), DFG (Germany) and the collaborating institutions in the Dark Energy Survey, which are Argonne Lab, UC Santa 
Cruz, University of Cambridge, CIEMAT-Madrid, University of Chicago, University College London, DES-Brazil Consortium, 
University of Edinburgh, ETH Zurich, Fermilab, University of Illinois, ICE (IEEC-CSIC), IFAE Barcelona, Lawrence Berkeley 
Lab, LMU Munchen and the associated Excellence Cluster Universe, University of Michigan, NOAO, University of Nottingham, 
Ohio State University, University of Pennsylvania, University of Portsmouth, SLAC National Lab, Stanford University, University 
of Sussex, and Texas A\&M University.

The G10/COSMOS redshift catalogue, photometric catalogue and cutout tool uses data acquired as part of the Cosmic Evolution 
Survey (COSMOS) project and spectra from observations made with ESO Telescopes at the La Silla or Paranal Observatories 
under programme ID 175.A-0839. The G10 cutout tool is hosted and maintained by funding from the International Centre for 
Radio Astronomy Research (ICRAR) at the University of Western Australia. Full details of the data, observation and catalogues 
can be found in \citet{Davies+2015} and \citet{Andrews+2017}, or on the G10/COSMOS website: cutout.icrar.org/G10/dataRelease.php

{\it Facilities:} $Blanco$ (DECam)

\end{document}